\documentclass[letterpaper,11pt]{article}
\usepackage[left=1in,right=1in,top=1in,bottom=1in]{geometry}
\usepackage[doublespacing]{setspace}
\usepackage{pdflscape}

\usepackage{soul,color}
\definecolor{seagreen}{RGB}{46, 139, 87}

\usepackage[reqno]{amsmath}
\usepackage[hidelinks]{hyperref}
\usepackage{amsthm}
\usepackage{amssymb,enumerate}
\theoremstyle{plain}

\usepackage{tikz}
\usetikzlibrary{fit,positioning,shapes,positioning,decorations.pathmorphing, decorations.pathreplacing,arrows}

\newcommand{\E}{{\rm I\kern-.3em E}}

\usepackage{framed}
\usepackage{caption}
\usepackage[authoryear,sort]{natbib}
\title{The origins of unpredictability in life trajectory prediction tasks\thanks{We thank the Fragile Families Challenge Board of Advisors for guidance. This study was supported by the Overdeck Education Research Innovation Fund, Russell Sage Foundation, NSF (1760052), and NICHD (P2-CHD047879). Funding for FFCWS was provided by the NICHD (R01-HD36916, R01-HD39135, R01-HD40421) and a consortium of private foundations, including the Robert Wood Johnson Foundation. For MJS, part of this work was done while he was the Infosys Member at the Institute for Advanced Study. Direct correspondence to Ian Lundberg, \href{mailto:ilundberg@cornell.edu}{ilundberg@cornell.edu}.\\{}\\
\textsuperscript{a}Department of Information Science, Cornell University\\
\textsuperscript{b}Department of Sociology, Princeton University\\
\textsuperscript{c}Department of Sociology, St. Joseph's University\\ 
\textsuperscript{d}Office of Population Research, Princeton University\\
\textsuperscript{e}Center for Information Technology Policy, Princeton University}}

\date{\today\vskip .05in {\normalsize Working paper. Comments welcome.}}

\author{Ian Lundberg\textsuperscript{a} \qquad Rachel Brown-Weinstock\textsuperscript{b} \qquad Susan Clampet-Lundquist\textsuperscript{c} \\ Sarah Pachman\textsuperscript{d} \qquad Timothy J. Nelson\textsuperscript{b} \qquad Vicki Yang\textsuperscript{b} \\ Kathryn Edin\textsuperscript{b} \qquad Matthew J. Salganik\textsuperscript{bde}}

\begin{document}

\maketitle

\begin{abstract}
\singlespacing
Why are life trajectories difficult to predict? We investigated this question through in-depth qualitative interviews with 40 families sampled from a multi-decade longitudinal study. Our sampling and interviewing process were informed by the earlier efforts of hundreds of researchers to predict life outcomes for participants in this study. The qualitative evidence we uncovered in these interviews combined with a well-known mathematical decomposition of prediction error helps us identify some origins of unpredictability and create a new conceptual framework. Our specific evidence and our more general framework suggest that unpredictability should be expected in many life trajectory prediction tasks, even in the presence of complex algorithms and large datasets. Our work also provides a foundation for future empirical and theoretical work on unpredictability in human lives.
\end{abstract}

\clearpage

\section*{Introduction}

Bella was born in a large American city around the year 2000.\footnote{All names in this paper are pseudonyms.} Bella's family was not wealthy by any means, but both of her parents graduated from high school, they were married soon after she was born, and both had stable employment. Bella's mom described Bella's childhood this way: ``She was nice and friendly, you know, just went to school and played and that was pretty much it.'' But by the time Bella turned 15, things looked very different. She was getting in fights at school and struggling in class. Eventually she dropped out of high school. Could Bella's transition from a happy childhood to struggling adolescence have been predicted? 

Questions like this about the predictability of human outcomes have been the subject of research and speculation at least since Cicero published \textit{On Divination} in 44 BCE \citep{vancreveld2020}. Although these questions usually seem intractable, in Bella's case we can be unusually confident in our answer. Bella was part of a multi-decade longitudinal social science study that collected detailed information about the life trajectories of thousands of families \citep{reichman2001fragile}. Then, hundreds of researchers used these data to create algorithms that attempted to accurately predict life outcomes \citep{salganik2020measuring}. Of all the algorithms trained on this rich dataset, the very best algorithm was not very accurate for Bella or overall. 

The goal of predicting someone's future might seem more rooted in science fiction than science, but life trajectory predictions are actually quite common: doctors predict the trajectory of patients, social workers predict the risk of mistreatment of children, landlords predict whether potential tenants will pay their rent, firms predict the productivity of potential employees, banks predict the creditworthiness of potential borrowers, and judges predict the likelihood that someone who was arrested will appear at trial.

While humans have historically carried out life trajectory predictions unaided, there is increasing interest in making life trajectory predictions using complex algorithms trained on large datasets. For example, \citet{kleinberg2018human} estimate that replacing judges with a machine learning model for decisions about bail could result in 40\% fewer people being subjected to pre-trial detention with no increase in crime and with a decrease in racial disparities. There are appropriate concerns about the fairness, accountability, transparency, ethics, and utility of these algorithms \citep{chouldechova2017fair,barocas2019,mitchell2021algorithmic,wang2023survey}, as well as cautious optimism that carefully designed algorithms might improve decisions and, by extension, well-being \citep{kleinberg2015prediction,kleinberg2018human}. 

Despite the fact that life trajectory predictions are common and often involved in high-stakes decisions, they have not been the focus of research about the life course (for a few exceptions, see \citealt{liou2023life}). This limited attention means that there is little scientific foundation for understanding the accuracy---and critically the inaccuracy---of these predictions. Nor is there an understanding of the fundamental processes that determine the predictability of life trajectories, and whether these might be overcome with more data, better algorithms, and improved theory.

This paper reports the results of qualitative interviews with many people like Bella and her mother. Because these interviews were sampled from a multi-decade longitudinal study in which hundreds of researchers built predictive algorithms, they offer an empirical approach to discovering the origins of unpredictability of life trajectory predictions. From these interviews, we inductively developed a conceptual framework based on a well-known mathematical decomposition.  The framework, which we illustrate with examples from the interviews, should help decision makers and guide future research.

\begin{figure}%[tbhp]
\centering
\includegraphics[width=.8\linewidth]{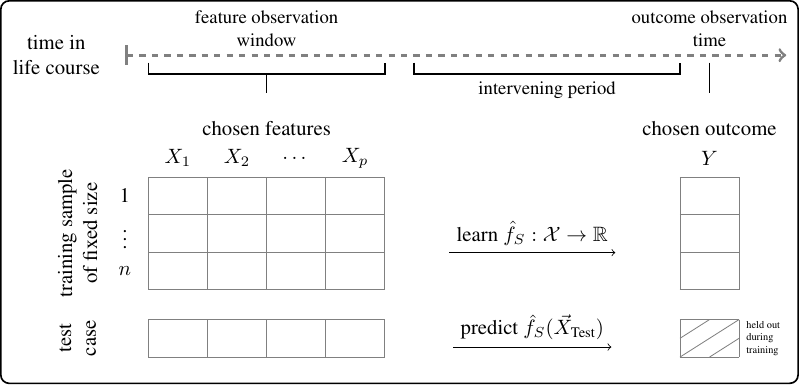}
\caption{Life trajectory prediction task. In the task we studied, the feature observation window is from the child's birth to age 9. The outcome observation time is age 15. The chosen features are 12,942 survey responses. The chosen outcome is self-reported Grade Point Average (GPA), which ranges from 1.00 (worst) to 4.00 (best).}
\label{fig:task}
\end{figure}

\section{Conceptual Framework}
\label{sec:framework}

A life trajectory prediction task is defined by three elements (Fig~\ref{fig:task}). The first element is a set of features (predictor variables) measured about a person. We refer to the time when features are measured as the feature observation window. In the task we studied, the features are a specific set of childhood experiences measured from birth to age 9. The second element of a life trajectory prediction task is an outcome variable, which we require to be measured at some point after the feature observation window. In the task we studied, the outcome is a particular measure of school performance at age 15. Between the feature observation window and outcome measurement is an intervening period, such as the 6 years between age 9 and 15. We refer to the length of the intervening period as the time horizon. The third element of life trajectory prediction task is the process that produces a training sample; this element includes both the sampling method and the sample size. An example is a simple random sample of a given size from a particular population.

Once a life trajectory prediction task is defined, people attempt the task using a learning approach: any procedure to create an algorithm that takes feature values as inputs and returns a predicted outcome. We conceptualize the learning approach broadly. A learning approach can be the process by which a human manually considers the observed data and draws on expertise to create a prediction rule. A learning approach can also be a machine learning procedure that automatically discovers a mapping from features to outcomes. A learning approach might combine human-driven and data-driven strategies. 

In this paper, we measure performance by mean squared out-of-sample prediction error. Mathematically, let $(\vec{X},Y)$ denote the feature vector and outcome for a random unit from the population, and let $\hat{f}_S$ denote a prediction function learned in the training sample $S$. Out-of-sample mean squared error is an estimate of expected squared error $\text{E}((\hat{f}_S(\vec{X}) - Y)^2)$ with expectation taken over an infinite population and over randomness in the training sample $S$. Future research could generalize our results to other performance metrics.

\begin{figure}
    \centering
    \includegraphics[width = .8\linewidth]{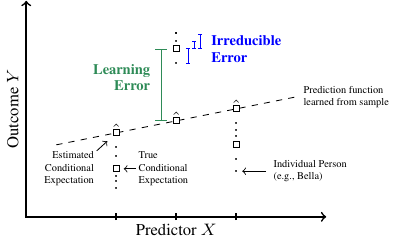}
    \caption{Origins of unpredictability: Irreducible error and learning error. See also \citet{berk2008statistical} Fig.~1.6. Each dot represents a person; people who share a single value on the predictor have many outcomes (each vertical set of dots). This type of error is fixed by the task definition, so we call it \textit{irreducible error}. Second, the estimated prediction function (regression line) is not equal to the true conditional expectation in each subgroup defined by $X$. The prediction is wrong in this case because the true relationship is not linear, but it could also be wrong due to sampling variability or bias. Because this component relates to the learning procedure, we call it \textit{learning error}.}
    \label{fig:origins}
\end{figure}

Each life trajectory prediction task can also be interpreted as an estimation task. For each person with feature vector value $\vec{x}$, consider the set of people who are observationally identical: the people who share this feature vector value. These people are identical from the perspective of an algorithm making predictions using these features, so their predicted values will be the same. The predicted value that would minimize expected squared error is the population average outcome within the group, $\text{E}(Y\mid\vec{X} = \vec{x})$. A prediction task with expected squared error loss is thus equivalent to an estimation task: estimate within-group mean outcomes. The estimation perspective clarifies that each individual prediction is inaccurate to the degree that one or both of the following is true: (1) the individual outcome is far from the within-group mean, and (2) the predicted value is far from the within-group mean (Fig~\ref{fig:origins}).

\begin{figure}[!htbp]
    \centering
    \includegraphics[width = .7\linewidth]{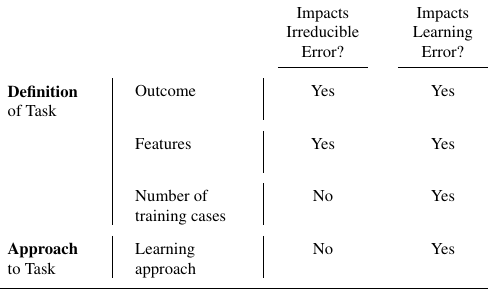}
    \caption{Inputs to irreducible error and learning error.}
    \label{fig:inputs}
\end{figure}

These two origins of unpredictability correspond formally to two components in a well-known mathematical, additive decomposition of expected squared error. The first component is the average within-group variance in individuals' outcomes. Because within-group variance is fixed by the features and outcome and does not involve the predicted values, we refer to this component as \textit{irreducible error}.  Irreducible error cannot be reduced by a new machine learning procedure; the only way it can be decreased is by changing the task. The second component of prediction error is the average squared difference between the estimated and true within-group mean outcomes. Because this component corresponds to errors in the learned prediction, we refer to this component as \textit{learning error}. Irreducible error and learning error additively comprise expected squared error (Eq~\ref{eq:decomposition} and Appendix Section S4) \citep{hastie2009elements}.

\begin{equation}
%\resizebox{.9\linewidth}{!}{
%\begin{tikzpicture}
%\node at (0,0) {$
\underbrace{
\text{E}\left(\left[Y - \hat{f}_{\mathcal{S}}\left(\vec{X}\right)\right]^2\right)
}_{\substack{\textbf{Prediction Error}\\\text{Expected squared}\\\text{prediction error}}}
=
\underbrace{
\text{E}\left(\text{V}\left[Y\mid\vec{X}\right]\right)
}_{\substack{\textbf{Irreducible Error}\\\text{Outcome variance}\\\text{given predictors}}}
+
\underbrace{
\text{E}\left(\left[\hat{f}_{\mathcal{S}}\left(\vec{X}\right) - \text{E}\left(Y\mid \vec{X}\right)\right]^2\right)
}_{\substack{\textbf{Learning Error}\\\text{Expected squared error}\\\text{for the conditional mean}}}
%$};
%\end{tikzpicture}
%}
\label{eq:decomposition}
\end{equation}
Some researchers further decompose learning error into model approximation and estimation error \citep{berk2008statistical}. Others further decompose learning error into bias and variance \citep{hastie2009elements}. We focus on irreducible and learning error because these two components have conceptually distinct sources: irreducible error is a function of the task only, whereas learning error is a function of both the task and the learning approach (Fig~\ref{fig:inputs}). We conceptualize the learning approach to include all decisions made by the researchers when going from the raw data to the final predictions. Our conceptual framework applies to any prediction function $\hat{f}_S$ no matter how it is created: any kind of statistical learning, human expertise, or combination of the two.

\section{Data}
\label{sec:data}

We took a mixed-methods approach \citep{small2011conduct} to discover the sources of irreducible and learning error in one life trajectory prediction task. Our team of 16 researchers conducted 114 semi-structured, qualitative interviews with 73 respondents in 40 families.

Every family we interviewed was part of the Future of Families and Child Wellbeing Study (FFCWS; formerly the Fragile Families and Child Wellbeing Study), a longitudinal study tracking the lives of thousands of families who gave birth around the year 2000 in 20 large U.S. cities \citep{reichman2001fragile}. Researchers collected survey data in five waves from the birth of the child through age 9, and then again in a sixth wave when children were 15 years old. The study gathered data from many respondents (child, child's parents, primary caregiver, teacher, etc.) on many different topics (material resources such as income, social factors such as parents' relationship, school characteristics, perceptions of the residential neighborhood, etc., Appendix, Fig. S3). Data also include psychometric testing of the child's cognitive development. Because of their depth, breadth, and quality, the FFCWS data have been used in more than 1,000 published papers \citep{ffpubs}.

To select a sample of families from FFCWS that would be especially informative about the origins of unpredictability, we drew on the results of the Fragile Families Challenge \citep{salganik2020measuring}. The Challenge was a scientific mass collaboration in which hundreds of researchers attempted to use the FFCWS data for six life trajectory prediction tasks. We focus on one of these tasks, in which researchers attempted to predict each child's average of self-reported grades in four subjects: English, history, math, and science. We refer to this outcome as grade point average (GPA), which can range from 1.00 (worst) to 4.00 (best). The 12,942 features were collected in the FFCWS from the birth of the child through age 9, including features such as family income, parental relationship status, and teacher reports of child behavior and school performance. The training set was 2,121 cases for which participants had access to the GPA at age 15. The task thus involved predictions over a six-year time horizon, from age 9 to age 15. Performance was evaluated on a holdout set of new observations by $R^2_\text{Holdout}$, which rescales out-of-sample mean squared error so that a score of zero corresponds to predicting the mean of the training data and a score of one corresponds to perfect prediction.  Despite using a rich dataset, a variety of theoretical approaches, and state-of-the-art machine learning, no researchers were able to make very accurate predictions: the best $R^2_\text{Holdout}$ when predicting GPA was 0.19 \citep{salganik2019, salganik2020measuring}. Predictability was also low for the other five prediction tasks with other outcomes.

The most accurate algorithm from the Challenge \citep{rigobon2019winning} provides a useful approximation for the best possible predictions for this task, given the expertise---substantive and methodological---available at that time. As such, children's outcomes that are not well-predicted by this algorithm may be particularly informative about the origins of unpredictability in this task. Therefore, we oversampled children whose GPAs were much higher than predicted and much lower than predicted. To avoid concerns about overfitting, we limited the sampling frame to children who were not in the training set. To capture the full distribution of predicted values, we stratified the sampling frame into terciles based on predicted GPA and conducted our sampling within terciles. To reduce the risk of misinterpreting the experiences of outliers, we also sampled a set of children whose actual and predicted GPA were similar. To increase our chance of observing structural forces that might be invisible to participants, we sampled children born in three different cities. Finally, to reduce the risk of motivated measurement, the primary interviewer for each case was not told the predicted and realized GPA. For additional information about the sampling and interview procedure, see Materials and Methods and Appendix, Sections S2 and S3. This design---which combines ideas from the qualitative and quantitative research traditions---was created to be informative about the origins of unpredictability.

\section{Results: Origins of Unpredictability}
\label{sec:results}

\subsection{Sources of Irreducible Error}

\begin{figure}
\includegraphics[width=\linewidth]{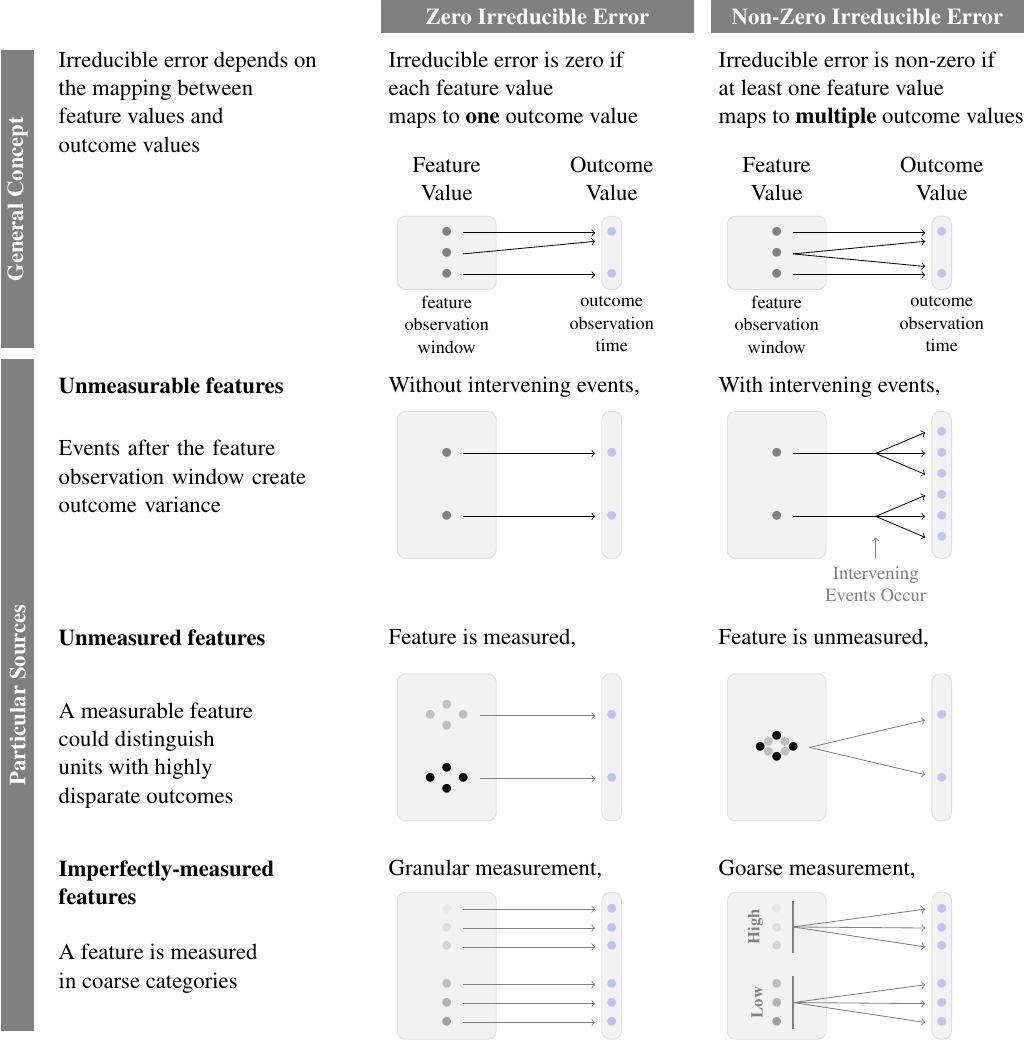}
    \caption{Sources of irreducible error.}
    \label{fig:irreducible}
\end{figure}

To the degree that observationally identical children have different outcomes, there exists irreducible error (Fig~\ref{fig:origins}). Here we focus on three non-exhaustive sources of irreducible error: unmeasurable features that occur after the feature observation window, unmeasured features that could have been measured because they occur during the intervening period, and imperfectly measured features (Fig.~\ref{fig:irreducible}). These three sources helped us organize many of the issues we learned about in our in-depth interviews, and we suspect that they will apply to some degree to any life trajectory prediction task.

\subsubsection{Unmeasurable features: Consequential intervening events}

In a life trajectory prediction, time elapses between measurement of predictors and realization of an outcome. Events in the intervening period of time cannot be measured and can create irreducible error. A consequential intervening event can upend a life trajectory in a long-lasting way, as when Bella's father died and her academic trajectory was thrown off course. A consequential event can also be more fleeting. Charles attended an online charter school and mostly performed well, working in the family dining room upstairs under parental supervision. But in the specific term for which GPA was measured, he attended school from the basement, where he often played online video games. That semester, Charles reported a 1.75 GPA, much lower than the predicted value of 3.15. Subsequently, his mother realized what was happening and said ``no more downstairs.'' His grades recovered: ``Ninth grade, I did terrible, then all the other years, I did As.'' For both Charles and Bella, an event occurred after all predictors were measured: Bella's father died, and Charles moved to school in the basement. These events may have caused their outcomes to differ from others who shared their feature sets. Yet the relevant events could not have been measured: they occurred during the period between predictors and outcomes. Consequential intervening events are an important source of irreducible error, particularly for life trajectory prediction tasks with long time horizons like that from age 9 to age 15.

\subsubsection{Unmeasured features}

Some features existed and could have been measured during the feature observation window, yet they are unmeasured. Features may be unmeasured for good reason, such as a survey designer facing a budget constraint. Yet unmeasured features can create irreducible error to the degree that they are independent of the measured predictors and relevant to the outcomes of many cases. Our qualitative interviews did not reveal a small set of additional predictors that we think would have greatly improved predictive performance. This is perhaps unsurprising---the FFCWS data included thousands of predictors collected with guidance from sociologists, psychologists, economists, and social work scholars. Yet, we did find examples of unmeasured predictors that seemed to be important in specific cases.

For example, Lola’s social network was particularly important. While her mother engaged in dangerous illegal activities, Lola got ready for school each day in the care of an elderly neighbor. Lola's grandparents provided health insurance and an address to enroll her in a better school, and ultimately remodeled their basement so that Lola and her mother could move in. In recent years, her mother was stably employed by an aunt in a family business. Perhaps if Lola's network had been measured, an algorithm could have better anticipated her 3.75 GPA, which outpaced the predicted value of 3.04.

\subsubsection{Imperfectly-measured features}

Sometimes a feature was measured during the feature observation window, but it was measured imperfectly. This imperfect measurement can create irreducible error. When considering imperfect measurements, researchers often focus on respondent misreporting \citep{biemer1991measurement}, but it can come from many other sources as well.  For example, imperfect measurement can also arise from limitations inherent to survey research when a continuous construct may be measured in coarsened categories.

For example, respondents at age 9 answered a question ``How close do you feel to your mom?'' with four response options from ``extremely close'' to ``not very close.'' Hennessey chose ``not very close.'' Her actual GPA of 1.25 was far below the predicted value of 2.71. One explanation for this poor prediction is coarse measurement: she needed an answer choice beyond ``not very close.'' In our qualitative interview, Hennessey reported that at times when she needed her mother, her mother ``blatantly ignored me.'' The two bickered and physically fought. Her mother sometimes kicked her out of the house or called the police. When asked directly if her stressful home life impacted her school performance, Hennessey noted that it ``affected me a lot.'' She recalled a particular incident when her mother told her that ``[y]ou better start treating me better, because I might not live that long.'' This warning was so frightening that she went to the principal's office because ``I couldn’t even focus in class...I was shaking. That was all I could think about. I was, like, crying in school, and they [school staff] had no idea what was wrong with me.'' Ultimately, Hennessey failed 8th grade and reported a low GPA in the FFCWS survey at age 15.

Hennessey's turbulent relationship with her mother was very consequential in her life, and it was only coarsely captured by the survey data. To a trained model, she appears the same as any other respondent whose relationship with their mother was ``not very close,'' even though hers was likely much worse than theirs. Imperfect feature measurement thus made it harder to predict Hennessey's outcome.

\subsection{Sources of Learning Error}

Learning error exists to the degree that predicted values $\hat{f}_S(\vec{x})$ are far from the unknown conditional mean $\text{E}(Y \mid \vec{X} = \vec{x})$ (Fig~\ref{fig:origins}). Here we focus on what makes learning error high in life trajectory prediction tasks, especially with survey data: these tasks are likely to involve many features with a limited number of cases and limited amounts of expert knowledge. These characteristics together make conditional means difficult to estimate.

Life trajectories are the consequence of many inputs. For this reason, tasks are likely to involve many features. Our case study task involved 12,942 features selected by domain experts for their relevance to the life course. Even if each feature were binary, the number of possible feature vectors would be $2^{12,942}$, substantially more than the number of atoms in the universe.\footnote{This claim assumes about $10^{80}$ protons in the universe \citep[p.~85]{barrow2002constants}, and there are fewer atoms than protons.} The impossibility of learning in such a space might suggest that tasks should be defined with fewer features. But concerns about irreducible error point the opposite direction: our interviews suggest that accurate prediction might actually require even more features. For example, one child told us about a wealthy out-of-state family who mentored him since they were connected through a program for urban youth when he was in middle school. Another told us about a landlord who took an interest in his family (the tenants) and voluntarily built a home gym in the basement so the youth could follow his passion for fitness. If we wanted irreducible error to be closer to zero, we might choose life trajectory prediction tasks with an even bigger number of features.

When there are many features, however, learning is possible only with a vast number of cases and / or a vast amount of expert knowledge outside of the data. The number of cases is limited by practical constraints; the costs of following people over time imply that longitudinal surveys typically involve only a few thousand people at most. In the absence of a vast number of cases, one could lean on expert knowledge. Perhaps an expert could somehow specify a small number of features---either in the original data or derived from the original data---that allow for accurate predictions? We see no evidence that such expert knowledge currently exists about the life course, and it is unclear whether it will ever exist.

Machine learning may seem to offer a way out: perhaps an expert could narrow the class of possible models so that the data could then choose the best among the candidates. Suppose an expert narrows the feature space from 12,942 to 1,000 features, and argues for a linear model with interactions involving no more than two variables at a time. But then there are 1,000 main effects and (1,000 choose 2) interaction terms for machine learning to choose among: a total of 500,500 parameters. It is unreasonable to expect machine learning to magically find that many of these parameters are truly zero and also estimate the non-zero ones precisely. Machine learning is certainly a step forward; it might yield a sparse approximation that is better than what an expert would produce alone. But the magnitude of learning error may still be high in an absolute sense.

\section{Discussion}
\label{sec:discussion}

This paper defined life trajectory prediction tasks and developed conceptual arguments that predictability will generally be low for these tasks because of two important origins: irreducible error and learning error. We conclude by speculating about generalization to other life trajectory prediction tasks and discussing implications for policy and for science.

\subsection{Generalizing to Other Life Trajectory Prediction Tasks}

\begin{figure}
    \centering
    \includegraphics[width = .6\linewidth]{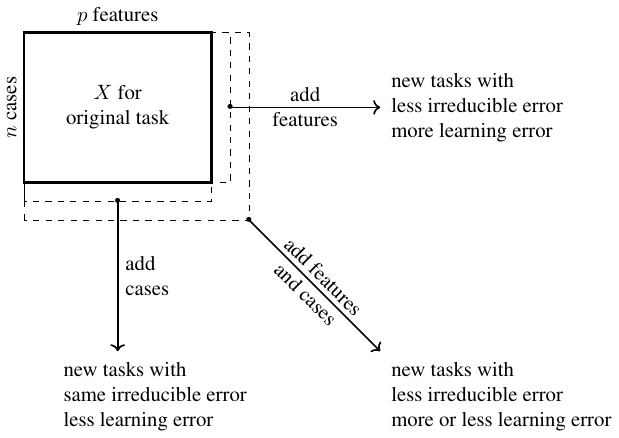}
    \caption{Generalizing to other life trajectory prediction tasks.}
    \label{fig:growth}
\end{figure}

Researchers may believe that high prediction error has an easy answer: more data! Unfortunately, the relationship between more data and prediction error is complex. More data could mean three different things---more cases, more predictors, or both---each with different implications for irreducible and learning error (Fig.~\ref{fig:growth}).

With more cases, learning error would be smaller but irreducible error would be unchanged. With more features, irreducible error would be smaller but learning error might increase because the number of unique feature values would grow exponentially with the number of features (the curse of dimensionality, \citealt{hastie2009elements}). With more cases and more features, prediction error might decrease or increase: irreducible error might decrease, but learning error might increase to the degree that the number of added cases is insufficient to learn accurately about the added features. In the extreme, there are deep questions about predictability in a world with infinite features \citep{li2021multi}. In practice, cost constraints limit the number of features and cases when life trajectory predictions are made with longitudinal survey data. Digital and administrative data held by companies and governments may offer qualitatively more cases and features. However, the features measured in digital and administrative data may be less useful for prediction than those measured in surveys \citep{salganik2018bit,bjerre2021task}.

Because we study only one task, we cannot draw firm conclusions about predictability in general. Yet our general conceptual framework and particular evidence leads us to speculate that for life trajectory prediction tasks using longitudinal survey data, low levels of predictability will be the norm. The inputs that create irreducible and learning error in our setting are likely to exist in many life trajectory prediction tasks. Two classes of tasks that might deviate from this pattern are: 1) tasks for which a natural low-dimensional representation maps predictors to outcomes, such as when a lagged outcome is a good predictor of that outcome in the future and 2) tasks for which the time horizon is very short (e.g. one day). However, we think that natural low-dimensional representations and short time horizons are the exception rather than the norm for life trajectory prediction tasks of interest in policy and science. Ultimately, our speculation requires empirical verification, refutation, and refinement. The strongest evidence will come from predictions pre-registered before the outcomes have taken place (e.g., \citealt{hegre2021can}) or from projects using the common task method (e.g., \citealt{salganik2020measuring}).

\subsection{Implications for policy and for science}

Much of the excitement about prediction for policy may stem from a belief that big data and machine learning magically lead to accurate predictions. We show instead that for life trajectory prediction tasks, there are deep reasons to expect unpredictability. Therefore, decision makers should re-orient their expectations and anticipate that life trajectory predictions---generated by humans or by algorithms---may be inaccurate. Further, decision makers should recognize that in many practical situations accurate prediction is a means to an end, not an end in itself \citep{liu2023}. In these cases, decision makers should focus less on accuracy and more on impact: the extent to which decisions informed by improved prediction actually lead to better outcomes \citep{murphy1993, sachs2020aim}. Unfortunately, the relationship between accuracy and impact can be complex and context dependent \citep{murphy1993, katz1987quality}. Impact evaluations for predictive models create important questions at the intersection of prediction, decision theory, and causal inference \citep{imai2023,wang2023survey}.

For many social scientists, individual-level prediction has not been a goal of research. Instead, a more common goal has been to describe difference between the mean outcome for different groups (e.g., difference in average life expectancy for people in different demographic groups) \citep{goldthorpe2016sociology}. Limits to predictability pose no direct threat to a focus on between-group mean variation; group means can be well-estimated even if the outcomes of individuals within each group vary substantially \citep{smith2011epidemiology, zhang2023illusion}. But if irreducible error is high, it would suggest an important complementary goal: describing within-group variability \citep{western2009}. Further, just as researchers currently seek to identify mechanisms that create between-groups differences, one can imagine a parallel search for processes that lead to within-group differences. A pivot from focusing on between-group variability to focusing on within-group variability could be a bridge between existing social science research traditions and research focused on limits to life trajectory prediction.

For researchers focused specifically on measuring and understanding the fundamental limits to predictability for life trajectory prediction tasks, additional empirical and theoretical work is clearly needed and can be guided by the definitions and conceptual framework developed in this paper. Empirically, researchers could quantitatively estimate prediction error for many tasks where one aspect of the task is systematically varied (e.g., the feature sets, outcomes, sample sizes) \citep{bjerre2021task,puterman2020predicting}. These empirical studies would be particularly valuable if they could estimate not just prediction error, but also irreducible error and learning error, which is possible in at least some settings \citep{fudenberg2022measuring}. Such studies should reveal which elements of a task are most central to prediction, irreducible, and learning error. Further, researchers could develop models that help reveal the social processes that lead of irreducible and learning error \citep{liou2023life} and models that yield sharp predictability bounds for specific data generating processes \citep{martin2016exploring}. This empirical and theoretical work on fundamental limits to life trajectory prediction may be informed by research about fundamental limits to prediction in other fields, such as the weather \citep{alley2019advances,bauer2015quiet} and in financial markets \citep{malkiel2003efficient}. Ultimately, this new research would lead to better frameworks for understanding life trajectories like Bella's. 

\section{Materials and Methods}
\label{sec:materials}

Our interviews were designed to study unpredictability. Each interview traced the life history of the young adult from birth through the time of the interview. The interview guide focused on three periods: 1) the feature observation window of birth to age 9, 2) the intervening period of age 9 to 15, and 3) age 15 to the time of the interview, which was after outcomes were measured. Appendix Section S6 and S7 provide the interview guides. Interviews were conducted in pairs, where only one of the two interviewers was aware of the outcome while conducting the interview. The interviewer who did not know the outcome conducted the interview. The interviewer who was aware of the outcome asked follow-up questions at the end. All interviews were recorded and transcribed. Appendix, Section S3 describes the interview procedure. Data collection was approved by the Princeton University IRB (\#10564).

We analyzed each interview inductively. Several members of the team independently answered a series of questions about each case and then met to discuss it. The themes in the paper emerged from these discussions. As themes crystallized, we switched to a format where a single researcher would write a case summary, other researchers would read the case summary and interview, and then we would meet to discuss and finalize the summary.

\begin{singlespacing}
\bibliography{bibliography}
\end{singlespacing}

\clearpage
\appendix
\setcounter{figure}{0}   
\setcounter{table}{0} 
\setcounter{section}{0} 
\renewcommand*\thetable{S\arabic{table}}  
\renewcommand*\thefigure{S\arabic{figure}}  
\renewcommand*\thesection{S\arabic{section}}  

\section*{APPENDIX}

\section{Acknowledgments}

We thank the following people who served as interviewers in this project: Bobbi Brashear, Rachel Brown-Weinstock, Maria Canals, Kristin Catena, Susan Clampet-Lundquist, Katie Donnelly, Kathryn Edin, Kaitlin Edin-Nelson, Alexus Fraser, Sarah Gold, Ashley Hyman, Ian Lundberg, Stefanie Mavronis, Timothy Nelson, Matthew Salganik, and Vicki Yang.

We thank the following people for feedback on the manuscript: Abdullah Almaatouq, Alyssa Battistoni, Jennie E. Brand, Elizabeth Bruch, Emily Cantrell, Kyla Chasalow, Siwei Cheng, Diag Davenport, Ben Edelman, Sayash Kapoor, Jennifer Lee, Jakob Mokander, Jonathan Murdoch, Daniel Rigobon, Brandon Stewart, Beza Taddess, Keyon Vafa, Akshay Venkatesh, and Haowen Zheng.

\section{Detailed Sample Information}

This project is embedded within the Future of Families and Child Wellbeing Study (FFCWS), a probability sample of children born in 1998–2000 in U.S. cities with populations over 200,000. The study is clustered in cities of birth. The Fragile Families Challenge assessed predictability of life outcomes in a subsample of 4,242 children born in 18 U.S. cities. As described by \citet{salganik2020measuring}, half of the sample (2,121 families) were provided to social and data scientists to build predictive models, one-eighth of the sample (530 families) were used to provide instant feedback on the predictive performance of submissions, and three-eights of the sample (1,591 families) were held out and used to evaluate predictive performance at the end of the Challenge.

\subsection{Sample Selection Process: Selecting Families for Qualitative Interviews}

The sample for qualitative interviews was drawn from among the 1,591 families in the holdout set. We focused on the children born in three FFCWS cities who were not missing the outcome variable (Grade Point Average, hereafter GPA) at age 15. Our sampling strategy sought to (1) assign non-zero sampling probability to every family, (2) ensure representation across cities and predicted GPA, and (3) ensure representation across residual GPA, with oversamples of those with GPAs much higher and much lower than expected.

\begin{figure}[!ht]
    \centering
    \includegraphics{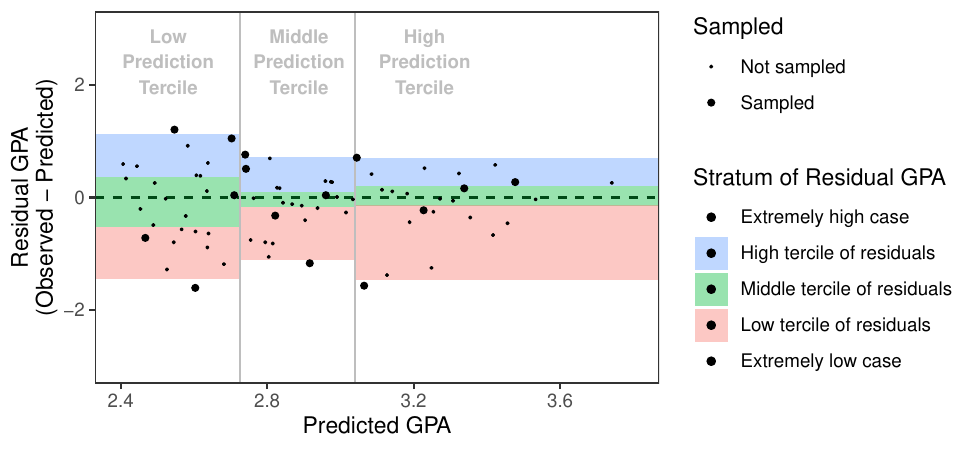}
    \caption{\textbf{Sample Selection Process.} The sample was stratified by city, predicted GPA, and residual GPA. Figure depicts the sampling process for one of the three cities; the two others were analogous. There were 3 strata of predicted GPAs: low, middle, and high. There were 5 strata of residual GPAs. The two extreme strata each contained only the one family with the most extreme positive or negative residual within the city $\times$ predicted GPA stratum; these families (6 per city, 18 in total) were sampled with probability 1. The three non-extreme strata were defined by terciles of residual GPA among the remaining cases within the city $\times$ predicted GPA categories. One family was sampled at random within each stratum (9 per city, 27 in total). Overall, the stratification procedure produces a sample that contains equal numbers of families from all three cities of low, middle, and high predicted GPA, with the most extreme residual GPAs oversampled but with cases selected across the range of residual GPA.}
    \label{fig:sample_selection_process}
\end{figure}

Fig.~\ref{fig:sample_selection_process} illustrates the resulting sampling strategy for one city. We stratified families into terciles within each city based on the GPA that was predicted by the most accurate submission to the Challenge. This produced three equally-sized strata: those with low, middle, and high predicted GPAs. Next, we stratified within each (city $\times$ predicted GPA) stratum by the residuals of the prediction: how much the actual GPA was better or worse than predicted. We sampled the families with the most positive and negative residuals with probability 1. We hoped these families would be especially informative for learning about unpredictability. We then partitioned the remaining observations into terciles of residual GPA: low, middle, and high. In each of these terciles, we randomly sampled 1 out of the 6-9 families.

Overall, this design yielded a full sample of 45 families with 15 per city (Fig.~\ref{fig:sample_selection_process}). The 15 include 5 each whose predicted outcomes are low, medium, and high. Each set of 5 within a stratum of predicted GPA includes the two respondents with the most unexpected outcomes and three respondents whose outcomes capture a range of the remaining residual values.

\subsection{Respondent Recruitment Process: Nonresponse, Refusals, and Replacements}

\begin{figure}[!ht]
    \includegraphics{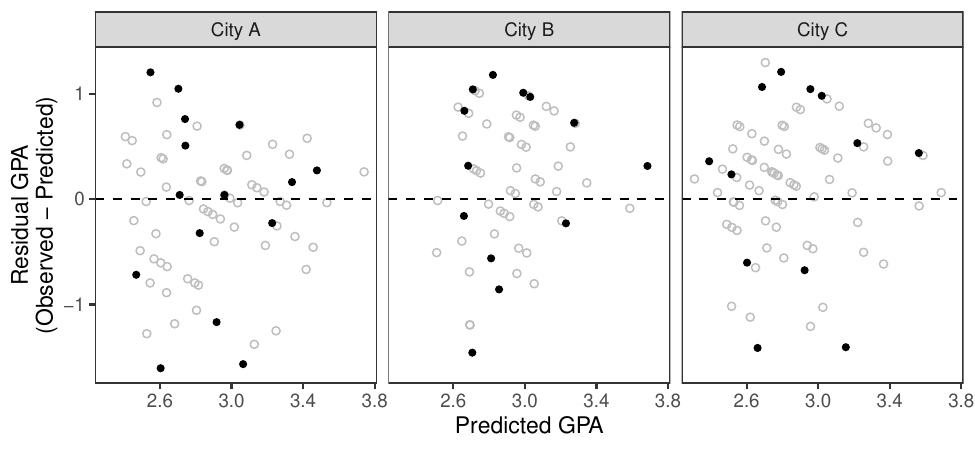}
    \caption{\textbf{Respondent Recruitment Process.} Responding families (solid dots) are distributed across the range of predicted and residual GPA, with an intentional oversample at extremely large positive and negative residuals.}
    \label{fig:respondent_recruitment_process}
\end{figure}

Fig.~\ref{fig:respondent_recruitment_process} shows the final set of responding families. We succeeded in speaking with at least one of the youth or primary caregiver in 40 families, of whom 38 were chosen as part of the initial sample selection process. The responding families in Fig.~\ref{fig:respondent_recruitment_process} are scattered around the sampling frame in a manner visually analogous to the sampled families in Fig.~\ref{fig:sample_selection_process}. 

For each sampled family, we made multiple contact attempts by U.S. mail, by phone, and when possible by email, social media, or via an in-person visit. Seven of the sampled families did not respond (Table \ref{tab:nonresponding_cases}). These non-responders can be placed into three groups. First, two families refused to participate: one case expressed hesitancy to participate at first and subsequently hung up immediately each time we called, 
and one case initially had difficulty scheduling and subsequently refused to participate. 
Second, for three families, either the youth or primary caregiver (or both) agreed to participate but we were unable to schedule an interview with them. Finally, for two families, we had difficulty contacting them: one case seemed to have correct contact information but did not yield a response, 
and the other had no up-to-date contact information. We presume the respondent never received our messages. 

When we could not reach a case or they declined to participate, we planned to replace that case with another similar case. Because of constraints on time and funding, we ultimately carried out this procedure for only 2 of the 7 families that were sampled but did not participate, as summarized in Table \ref{tab:nonresponding_cases}. 

\subsection{Differential Nonresponse}

If the responding and nonresponding families differ systematically along a variable relevant to the study, differential nonresponse could produce misleading conclusions. To assess differential nonresponse, one can compare the sampled respondents (Fig.~\ref{fig:sample_selection_process}) with the final respondents (Fig.~\ref{fig:respondent_recruitment_process}). In City A, we spoke with all 15 sampled families, so there was no differential nonresponse. The replacement case in each of cities B and C had similar predicted and residual GPA to the originally sampled case. While these replacements do not resolve differential nonresponse along unobserved variables, they are similar on observed variables. The greatest threats of differential nonresponse come from the 5 non-responding families who were not replaced (see Table \ref{tab:nonresponding_cases}). Two were in City B and 3 were in City C. Of the non-replaced families, 3 fell in the middle tercile of predicted GPA and 2 fell in the upper tercile of predicted GPA within their cities. Examining the residual categories of each non-replaced family within its (city $\times$ predicted GPA) stratum, 1 had the most negative residual within its stratum, 2 were in the bottom third of non-extreme residuals within their strata, and 2 were in the middle third of non-extreme residuals within their strata. This provides some evidence that our respondents under-represents those whose GPAs were lower than their predicted values. However, the relative rarity of the non-replaced families compared to the sampled and completed families suggests that differential non-response may only have minor implications for our conclusions.

\section{Interview Procedure}

A team of 16 researchers conducted 114 interviews with 73 respondents. The interviews were with both young adults and their primary caregivers: 66 interviews were with 39 young adult respondents and 48 interviews were with 34 primary caregiver respondents. There were more interviews than respondents for two reasons: 1) we planned to conduct two interviews with each young adult, and 2) some interviews were interrupted and needed to be rescheduled. All interviews were conducted by a pair of researchers, and most interviews were conducted in person, although some were conducted by phone when the respondent was geographically distant or preferred to speak by phone. We conducted interviews in both English and Spanish.

Our interview guide was designed to elicit life histories from youth and retrospective accounts about the youth’s life experiences from the primary caregivers, with a special focus on the time between when the young adult was 9 and 15 years old. The interview guides are included in Appendix Section S6 (Young Adult) and S7 (Primary Caregiver). 

When designing the interview protocols for this study, we had to decide how much prior information about the young adult and family to make available to the interviewers. In particular, we considered whether the interviewer should know the prediction error for the young adult's GPA. On one hand, knowing the error might allow the interviewer to probe responses that might help us better uncover the factors that are leading to unexpected outcomes. On the other hand, if the interviewer probed differently based on the outcome, we risked a circular design that could be used to justify known outcomes but would not help us understand other families with unknown outcomes. Ultimately, we decided on a hybrid design that provided us the benefits of both.

Both interviewers had access to basic information about the family: dates of past interviews, the age of the youth at those interviews, and information about the youth’s contact with their biological parents (e.g. whether resident or nonresident) at the last interview wave. Critically, however, the primary interviewer was unaware of the age 15 information and the GPA residual but the secondary interviewer was not. For the first part of the visit, the primary interviewer conducted the interview. At the end, the primary interviewer turned to the secondary interviewer and asked if there were any more questions. Then the secondary interviewer---who was already aware of the outcome during the entire interview---was able to probe responses that were particularly interesting given the young adult’s residual and survey responses at age 15.

\clearpage
\section{Decomposing Prediction Error: Mathematical Derivation}

The proof below derives the decomposition presented in Eq (1) in the main text. When writing expectation and variance operators $E()$ and $V()$, we include subscripts on these operators to define the random variables over which expectation and variance is being taken.

\begin{align}
&\text{E}_{Y,\vec{X},S}\left[\left(Y - \hat{f}_S(\vec{X})\right)^2\right]\\
&\text{Next add 0} \nonumber \\
&=
\text{E}_{Y,\vec{X},S}\left[\left(Y - \text{E}_Y(Y\mid\vec{X}) + \text{E}_Y(Y\mid\vec{X}) - \hat{f}_S(\vec{X})\right)^2\right]\\
&\text{Next distribute} \nonumber \\
&=
\text{E}_{Y,\vec{X},S}\left[\left(Y - \text{E}_Y(Y\mid\vec{X})\right)^2\right] + \text{E}_{Y,\vec{X},S}\left[\left(\text{E}_Y(Y\mid\vec{X}) - \hat{f}_S(\vec{X})\right)^2\right] \nonumber \\
&\qquad + 2\text{E}_{Y,\vec{X},S}\left[\left(Y - \text{E}_Y(Y\mid\vec{X})\right)\left(\text{E}_Y(Y\mid\vec{X}) - \hat{f}_S(\vec{X})\right)\right] \\
&\text{Apply the law of iterated expectation to the second line} \nonumber \\
&=
\text{E}_{Y,\vec{X},S}\left[\left(Y - \text{E}_Y(Y\mid\vec{X})\right)^2\right] + \text{E}_{Y,\vec{X},S}\left[\left(\text{E}_Y(Y\mid\vec{X}) - \hat{f}_S(\vec{X})\right)^2\right] \nonumber \\
&\qquad + 2\text{E}_{\vec{X},S}\left(\text{E}_{Y}\left[\left(Y - \text{E}_{Y}(Y\mid\vec{X})\right)\left(\text{E}_Y(Y\mid\vec{X}) - \hat{f}_S(\vec{X})\right)\mid \vec{X},S\right]\right) \\
&\text{Next, note that in the inner expectation all terms are constant except $Y$.} \nonumber \\
&\text{Rewrite with the inner expectation applying only to $Y$.} \nonumber \\
&=
\text{E}_{Y,\vec{X},S}\left[\left(Y - \text{E}_Y(Y\mid\vec{X})\right)^2\right] + \text{E}_{Y,\vec{X},S}\left[\left(\text{E}_Y(Y\mid\vec{X}) - \hat{f}_S(\vec{X})\right)^2\right] \nonumber \\
&\qquad + 2\text{E}_{\vec{X},S}\left[\left(\text{E}_Y(Y\mid \vec{X},S) - \text{E}_{Y}(Y\mid\vec{X})\right)\left(\text{E}_Y(Y\mid\vec{X}) - \hat{f}_S(\vec{X})\right)\right] \\
&\text{Because sample is random, }\text{E}_Y(Y\mid\vec{X},S) = \text{E}_Y(Y\mid \vec{X}) \nonumber \\
&=
\text{E}_{Y,\vec{X},S}\left[\left(Y - \text{E}_Y(Y\mid\vec{X})\right)^2\right] + \text{E}_{Y,\vec{X},S}\left[\left(\text{E}_Y(Y\mid\vec{X}) - \hat{f}_S(\vec{X})\right)^2\right] \nonumber \\
&\qquad + 2\text{E}_{\vec{X},S}\bigg[\underbrace{\left(\text{E}_Y(Y\mid \vec{X}) - \text{E}_{Y}(Y\mid\vec{X})\right)}_{=0}\left(\text{E}_Y(Y\mid\vec{X}) - \hat{f}_S(\vec{X})\right)\bigg] \\
&= \underbrace{\text{E}_{Y,\vec{X},S}\left[\left(Y - \text{E}_Y(Y\mid\vec{X})\right)^2\right]}_\text{Irreducible Error} + \underbrace{\text{E}_{Y,\vec{X},S}\left[\left(\text{E}_Y(Y\mid\vec{X}) - \hat{f}_S(\vec{X})\right)^2\right]}_\text{Learning Error}
\end{align}

\clearpage

\section{Supplemental Figures and Tables}

\begin{table}
    \caption{\textbf{Nonresponding Cases and Replacements.} Seven families in the sample selection process did not respond, and two of these families were replaced with another case.}
\centering
\begin{tabular}{llll}
  \hline
    Status
    & Stratum 
    & Replacement Status 
    & Replacement Stratum \\ 
  \hline
    \\
    Non-contact
    & \begin{minipage}[t]{1.2in}
        City C\\
        High predicted GPA\\
        Low residual \\
    \end{minipage} &
    Not replaced \\
    \\
    %%%%%%%%%%%%%%%%%%%%
    Refusal
    & \begin{minipage}[t]{1.2in}
        City B \\
        High predicted GPA \\
        High residual
    \end{minipage}
    & Replaced
    & \begin{minipage}[t]{1.2in}
        City B \\
        High predicted GPA \\
        High residual
    \end{minipage} \\ 
    \\
    %%%%%%%%%%%%%%
   Never scheduled
   & \begin{minipage}[t]{1.2in}
       City C \\
       Middle predicted GPA \\
       Low residual
   \end{minipage}
   & Not replaced \\
    \\
   %%%%%%%%%%%%%%
    Never scheduled
    & \begin{minipage}[t]{1.2in}
        City C \\
        Middle predicted GPA\\
        Extremely high residual
    \end{minipage}
    & Replaced 
    & \begin{minipage}[t]{1.2in}
        City C \\
        Middle predicted GPA\\
        High residual
    \end{minipage}\\
    \\
   %%%%%%%%%%%%%%
    Refusal
    & \begin{minipage}[t]{1.2in}
        City B \\
        Middle predicted GPA\\
        Middle residual
    \end{minipage}
    & Not replaced \\ 
    \\
   %%%%%%%%%%%%%%
    Non-contact
    & \begin{minipage}[t]{1.2in}
        City B \\
        High predicted GPA \\
        Extremely low residual
    \end{minipage}
    & Not replaced \\ 
    \\
   %%%%%%%%%%%%%%
    Never scheduled
    & \begin{minipage}[t]{1.2in}
        City C \\
        Middle predicted GPA\\
        Middle residual
    \end{minipage}
    & Not replaced \\ 
    \\
    \hline
\end{tabular}
    \label{tab:nonresponding_cases}
\end{table}

\begin{figure}[!htbp]
    \includegraphics[width = \textwidth]{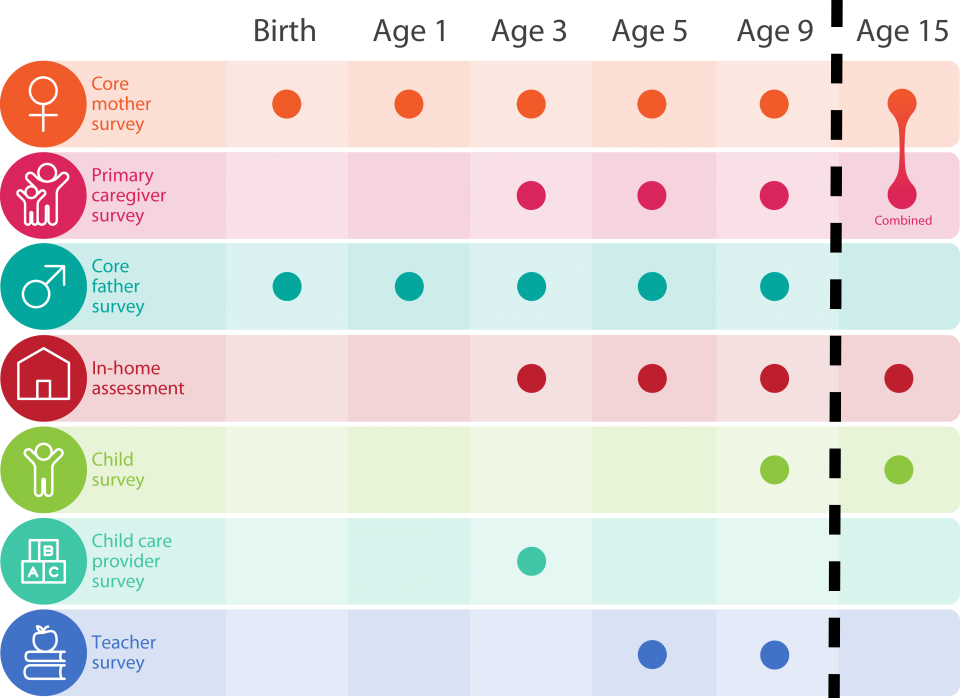}
    \caption{\textbf{Domains of predictors in the Fragile Families Challenge.} Image is from the Future of Families and Child Wellbeing Study: \url{https://ffcws.princeton.edu/}.}
    \label{fig:domains}
\end{figure}

\section{Interview Guide: Young Adult Interview}

\foreach \i in {1,...,16} {
\begin{tikzpicture}
    \node[draw] (page) at (0,0) {\includegraphics[page=\i, scale = .75]{interview_guides/interview_guide_youth.pdf}};
    \node[anchor = south] at (page.north) {Young Adult Interview Guide};
\end{tikzpicture}
\clearpage
}

\section{Interview Guide: Parent Interview}

\foreach \i in {1,...,10} {
\begin{tikzpicture}
    \node[draw] (page) at (0,0) {\includegraphics[page=\i, scale = .75]{interview_guides/interview_guide_parents.pdf}};
    \node[anchor = south] at (page.north) {Primary Caregiver Interview Guide};
\end{tikzpicture}
\clearpage
}

\end{document}